\documentclass[11pt,a4paper]{scrartcl}

\usepackage{ILD}

\usepackage[symbol]{footmisc}
\usepackage{feynmf}
\usepackage{textpos}
\usepackage{lineno}
\usepackage{wrapfig}

\title{Higgs self-coupling measurement at future $e^+e^-$ colliders}

\date{\today}
\addauthor{Julie Munch Torndal}{\institute{1,2}}
\addauthor{Jenny List}{\institute{1}}
\addauthor{Dimitrios Ntounis}{\institute{3}}
\addauthor{Caterina Vernieri}{\institute{3}}

\addinstitute{1}{Deutsches Elektronen-Synchrotron DESY, Notkestr. 85, 22607 Hamburg, Germany}
\addinstitute{2}{Department of Physics, Universit\"at Hamburg, Jungiusstra\ss e 9, 20355 Hamburg, Germany}
\addinstitute{3}{SLAC National Accelerator Laboratory, 2575 Sand Hill Road, Menlo Park, California, United States}

\abstract{The Higgs mechanism is a central part of the Standard Model which has not yet been fully established experimentally without the measurement of the Higgs self-coupling. Future linear $e^+e^-$ colliders are able to access centre-of-mass energies of 500 GeV and beyond and can therefore probe the Higgs self-coupling directly through the measurement of double Higgs production. A new analysis of the capability to measure the double Higgs-strahlung, $e^+e^-\to ZHH$, at a centre-of-mass energy of $500$\,GeV is ongoing based on the detailed, Geant4-based simulation of the ILD detector concept. This study has identified several aspects concerning the reconstruction techniques to fully exploit the detector potential, which are expected to improve precision reach and will be presented in this contribution. Additionally, the requirements that the Higgs self-coupling measurement puts on the choice of centre-of-mass energy will be evaluated as this is important for shaping the landscape of future colliders such as ILC or $C^3$.}

\titlecomment{Talk presented at the European Physical Society Conference on High Energy Physics (EPS-HEP2023), 21-25 August 2023, Universit\"at Hamburg, Hamburg, Germany.}

\graphicspath{ {.} }

\begin{document}

% generates the title page
\titlepage

% include source for sections
\section{Introduction}
Measuring the Higgs self-coupling is an important ingredient for reconstructing the Higgs potential and thereby establishing the Higgs mechanism experimentally. The value of the Higgs self-coupling can be probed either indirectly through loop-order corrections to the single-Higgs production cross-section, or by measuring the cross section of double Higgs production, sensitive to the self-coupling at tree-level. In either case, global interpretations combined with other Higgs, top and electroweak measurements, for instance in SMEFT, are required.
Single-Higgs measurements at two different centre-of-mass energies need to be combined in order to resolve ambiguities with other SMEFT operators, in particular $c_H$, c.f. Sec.\,12.5 of~\cite{ILCInternationalDevelopmentTeam:2022izu}, while sufficient collision energy, at or above $500$\,GeV is needed to produce di-Higgs events at future $e^+e^-$ colliders.  The two main production processes proceed either via di-Higgs strahlung $e^+e^- \to ZHH$, dominant for $\sqrt{s}<1$\,TeV or via $WW$ fusion $e^+e^- \to \nu_e\bar{\nu}_e HH$, dominant above $1$\,TeV.

\section{Measuring the SM Higgs self-coupling at Future $e^+e^-$ Colliders}\label{sec:lambdameasurement}
The possibilities to constrain the self-coupling from single-Higgs measurements have been studied at the time of the last update of the European Strategy for Particle Physics based on a SMEFT fit with 30 parameters assuming neutral diagonality~\cite{deBlas:2019rxi, narain2023future} combined with HL-LHC projections, in particular a projected precision of 50\% on the self-coupling from HL-LHC. This resulted in precisions of 49\% for FCCee at 240\,GeV and likewise for ILC at 250\,GeV.  Combining with higher energy runs, 33\% were obtained for FCCee at 365\,GeV, while ILC at 500\,GeV would reach 38\% from single-Higgs measurements alone.

However, future linear $e^+e^-$ colliders also offer tree-level access to the self-coupling via di-Higgs production. A precision of 36\% is a expected for CLIC1500 and 9\% for CLIC1500+3000~\cite{deBlas:2019rxi}, based on detailed simulation studies in~\cite{Roloff:2019crr}. For ILC, the last round of projections were performed between 7-10 years ago~\cite{Duerig:310520} at the proposed $E_{CM}$ of $500$\,GeV based on full simulation of the International Large Detector~\cite{behnke2013international, theildcollaboration2020international}. Assuming the SM predicted value for $\lambda$, the analysis found an expected precision of $26.6\%$ on $\Delta\lambda_{SM}/\lambda_{SM}$ based on 4 ab$^{-1}$ at $500$\,GeV combining the channels with $HH\rightarrow bbbb$ and $HH\rightarrow bbWW$. Including an additional run at $1$\,TeV, where the $\nu\nu HH$ process can be accessed, the precision improves to the $10\%$ level. 

With $\delta \lambda /\lambda \le 20\%$, a $5\,\sigma$ discovery can be claimed hence, the discovery potential for the Higgs self-coupling has been clearly demonstrated for CLIC3000 and ILC1000 already. Since these studies were done, a lot of progress has been made on some of the most relevant reconstruction tools. It is estimated that once propagated through the analyses, the self-coupling measurement could improve to better than $20\%$ on $\Delta\lambda_{SM}/\lambda_{SM}$ from the $ZHH$ measurement at 500\,GeV alone. A new analysis, likewise based on full, Geant4-based simulation of ILD, is underway to substantiate these estimates as well as to study the impact of the chosen centre-of-mass energy. These aspects will be discussed in the following section, before turning to the beyond-the-SM case.

\section{Choice of centre-of-mass energy and reconstruction performance} \label{sec:ecm}
After an initial run at the $ZH$ threshold, ILC and $C^3$ propose to run at $500$\,GeV and $550$\,GeV, respectively, targeting the direct $ZHH$ measurement as well as a full $CP$ analysis of the top quark's electroweak couplings and the top Yukawa coupling. At this point, it is not clear which energy in the range of $500$\,GeV to $600$\,GeV would be optimal for measuring the Higgs self-coupling: At higher energies, more $ZHH$ events are produced, and with more energy, the jets will be more boosted which might reduce misclustering and improve the jet pairing as well as the $b$-tagging performance, and we might see better kinematic separation of the signal and background. However, as $E_{CM}$ increases, contributions to the cross section from Feynman diagrams not containing the Higgs self-coupling grow faster, i.e.\ the relative sensitivity to $\lambda_{SM}$ decreases~\cite{Duerig:310520}.
\begin{figure}
    \centering
\begin{subfigure}{.3\textwidth}
  \centering
  \includegraphics[width=\textwidth]{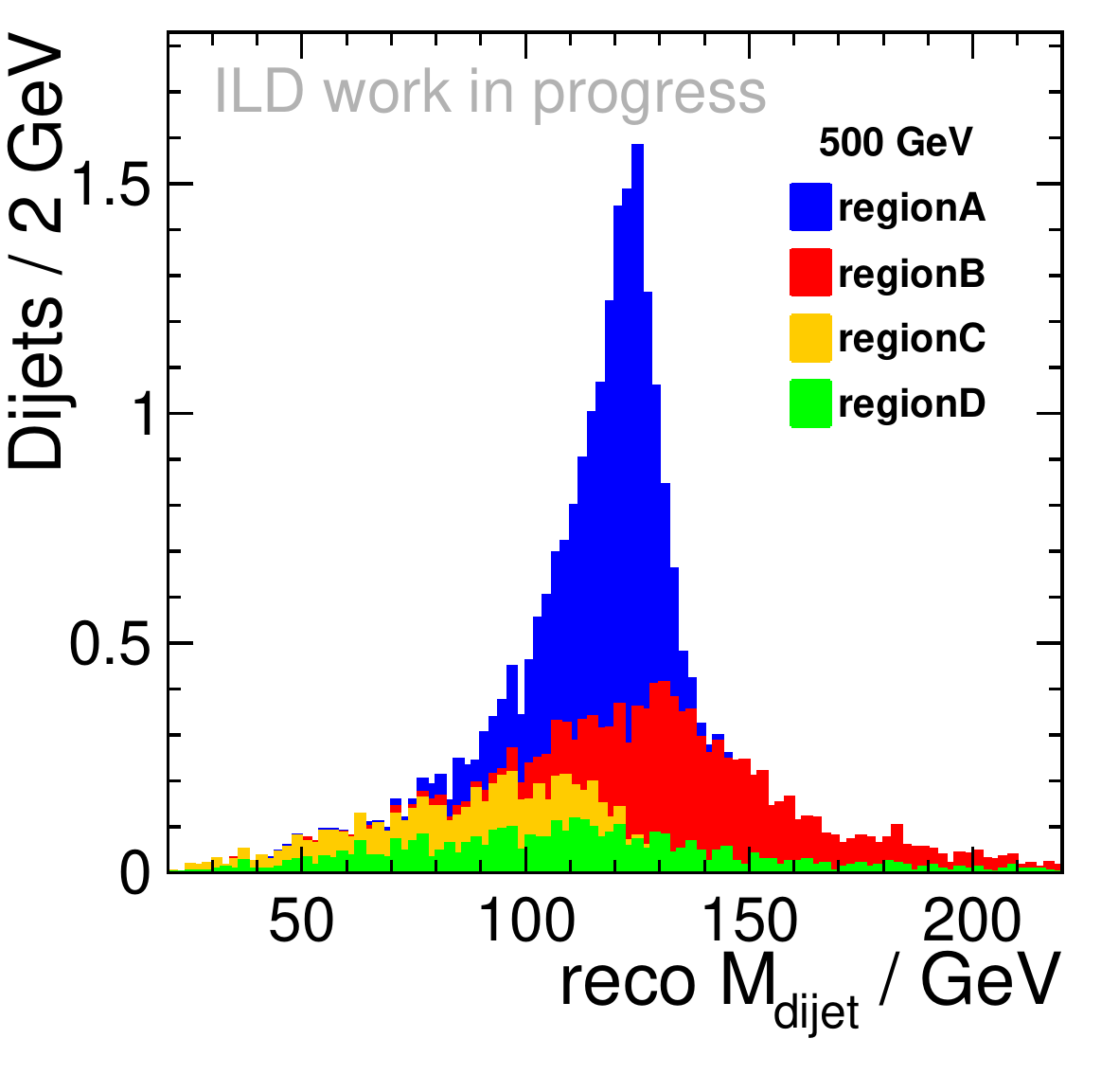}
  \subcaption{$500$\,GeV \label{fig:misclusteringdijetsmasses500}}
\end{subfigure}
\begin{subfigure}{.3\textwidth}
  \centering
  \includegraphics[width=\textwidth]{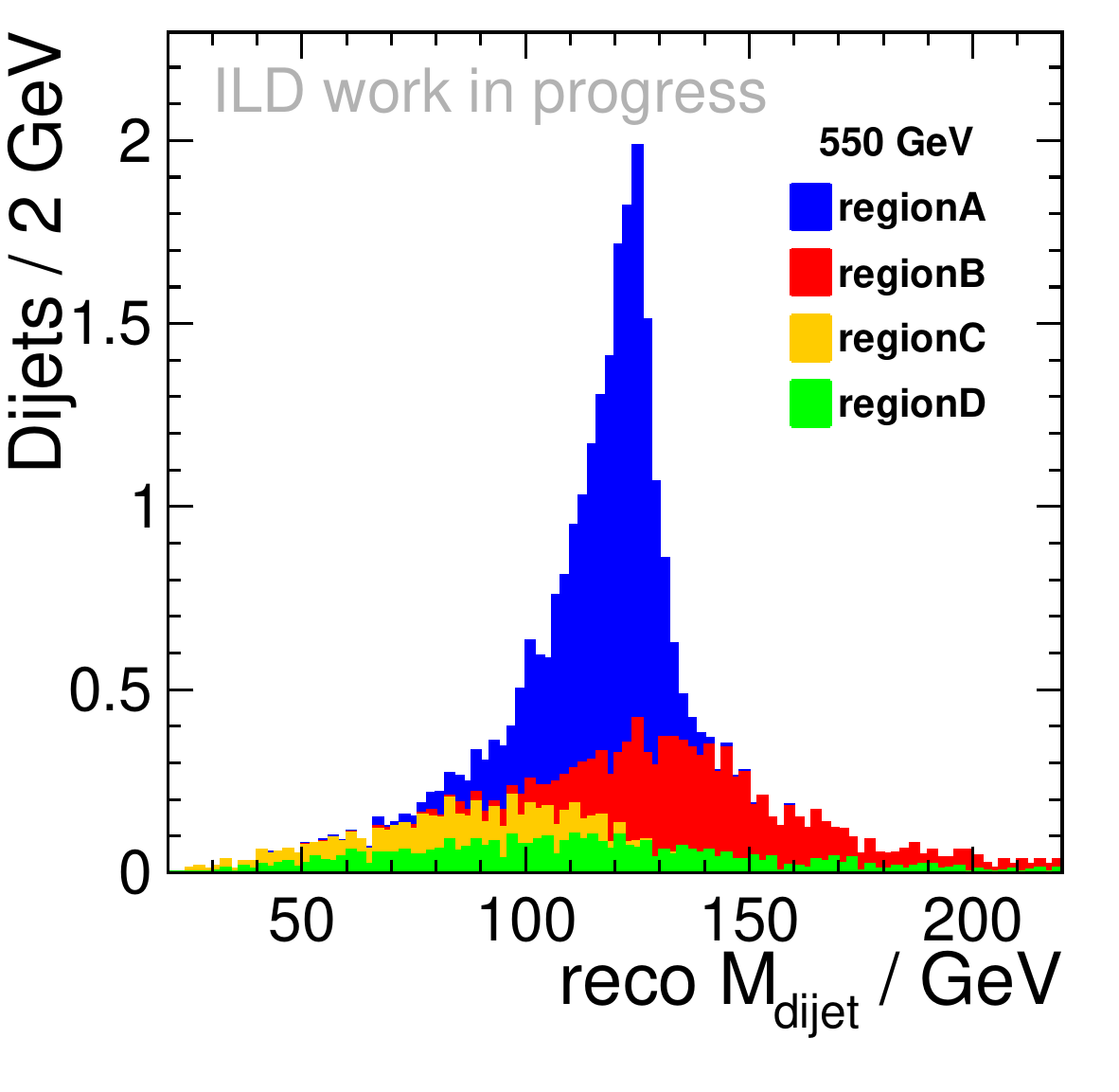}
  \subcaption{$550$\,GeV \label{fig:misclusteringdijetsmasses550}}
\end{subfigure}
\begin{subfigure}{.3\textwidth}
  \centering
  \includegraphics[width=\textwidth]{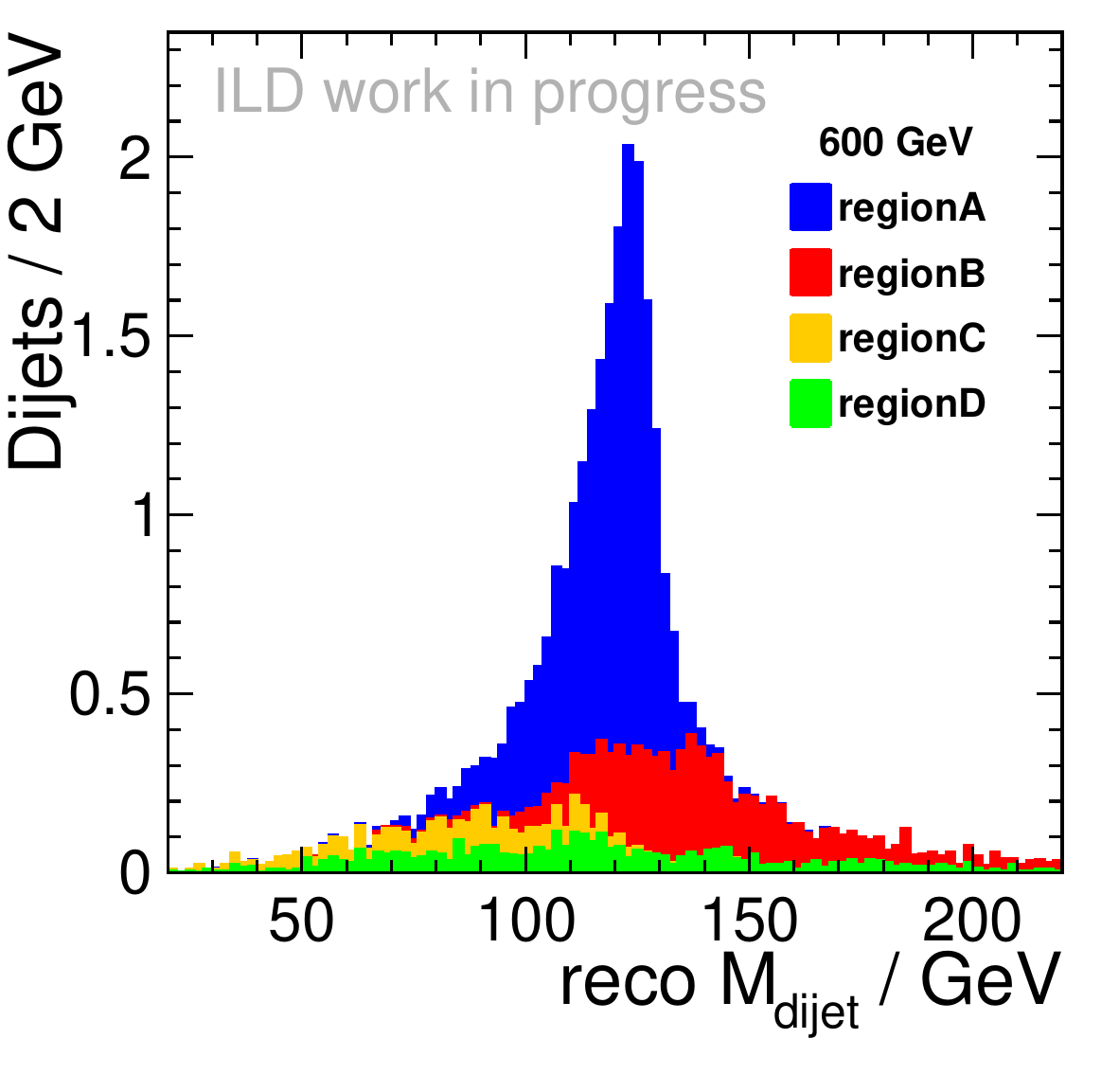}
  \caption{$600$\,GeV \label{fig:misclusteringdijetsmasses600}}
\end{subfigure}
\caption{Reconstructed di-jet mass distributions categorised by the jet clustering performance in terms of purity $\pi$ and efficiency $\epsilon$ at three different values of $E_{CM}$. A: $\pi$, $\epsilon > 0.95$; B: $\pi < 0.95$, $\epsilon > 0.95$, B: $\pi > 0.95$, $\epsilon < 0.95$; D: $\pi$, $\epsilon < 0.95$. } 
\label{fig:misclusteringdijetsmasses}
\end{figure}

$ZHH$ events have high jet multiplicities and at $E_{CM} \simeq 500$\,GeV the bosons are produced nearly at rest, causing jet-finding ambiguities that were found to reduce the sensitivity by almost a factor $2$ in the previous study~\cite{Duerig:310520}. In order to assess the state-of-the-art situation and develop mitigations, the di-jets are classified in terms of efficiency $\epsilon$ and purity $\pi$, defined as energy fraction of the true jet constituents contained in the reconstructed jet and the energy fraction of the particle flow objects in the reconstructed jet stemming from the true jet, respectively~\cite{torndal2023higgs}.  Figure~\ref{fig:misclusteringdijetsmasses} shows the reconstructed di-jet mass distributions in the channel with $ZHH\rightarrow \ell\ell bbbb$ at $500$, $550$ and $600$\,GeV separated into different misclustering classes. Di-jets in category A ($\epsilon$, $\pi > 0.95$) peak around the nominal Higgs mass, whereas tails are accounted for by the di-jets in categories B, C, and D. This shows that the tails are dominated by misclustering and not by the detector resolution. As the $E_{CM}$ increases, the fraction of di-jets in category A increases from 45.5\% at $500$\,GeV to 50.5\% at $550$\,GeV and 53.7\% at $600$\,GeV.

%\begin{wrapfigure}[12]{r}{0.48\textwidth}
\begin{figure}[htb]
    \centering
    \includegraphics[width=0.6\textwidth]{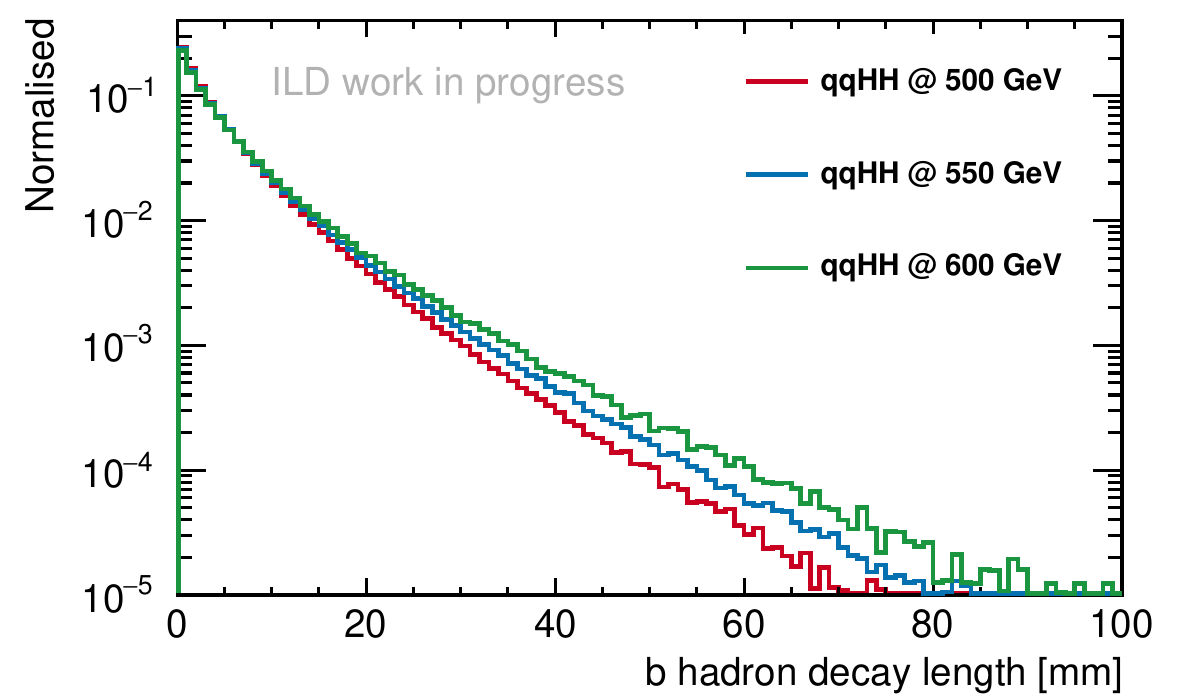}
    \caption{Decay lengths of weakly decaying  $b$-hadrons, comparing $500$\,GeV, $550$\,GeV and $600$\,GeV.}
    \label{fig:bdecaylengths}
\end{figure}    
%\end{wrapfigure} 
Since the last analysis was performed, the flavor tagging algorithm, LCFIPlus~\cite{Suehara:2015ura}, has seen substantial improvements. A 5\% relative improvement in the $b$-tagging efficiency (at the same background rejection rate) has been achieved already a few years ago~\cite{Taikan_presentation} and could lead to an 11\% relative improvement in the self-coupling precision~\cite{Duerig:310520, torndal2023higgs}. Figure~\ref{fig:bdecaylengths} shows that the decay lengths of the weakly decaying $b$-hadrons increase visibly with $E_{CM}$, as expected from the increasing boost. The $b$-hadron decay lengths are important input to separate $b$-jets from $c$- and light flavor jets, but a quantitative assessment of the effect on the $b$-tagging is work-in-progress.

\begin{figure}
\centering
\begin{subfigure}{.495\textwidth}
  \centering
  \includegraphics[width=0.95\textwidth]{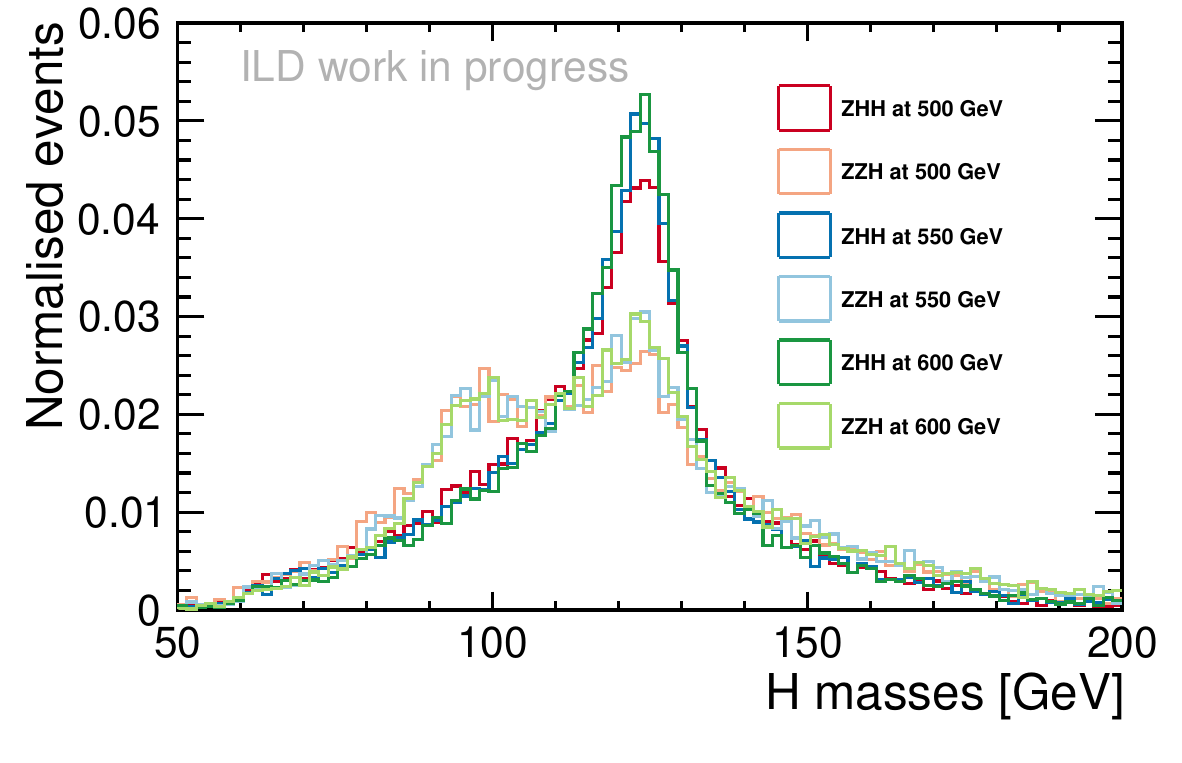}
  \subcaption{$ZHH$ hypothesis \label{fig:jetpairingZHH}}
\end{subfigure}
\begin{subfigure}{.495\textwidth}
  \centering
  \includegraphics[width=0.95\textwidth]{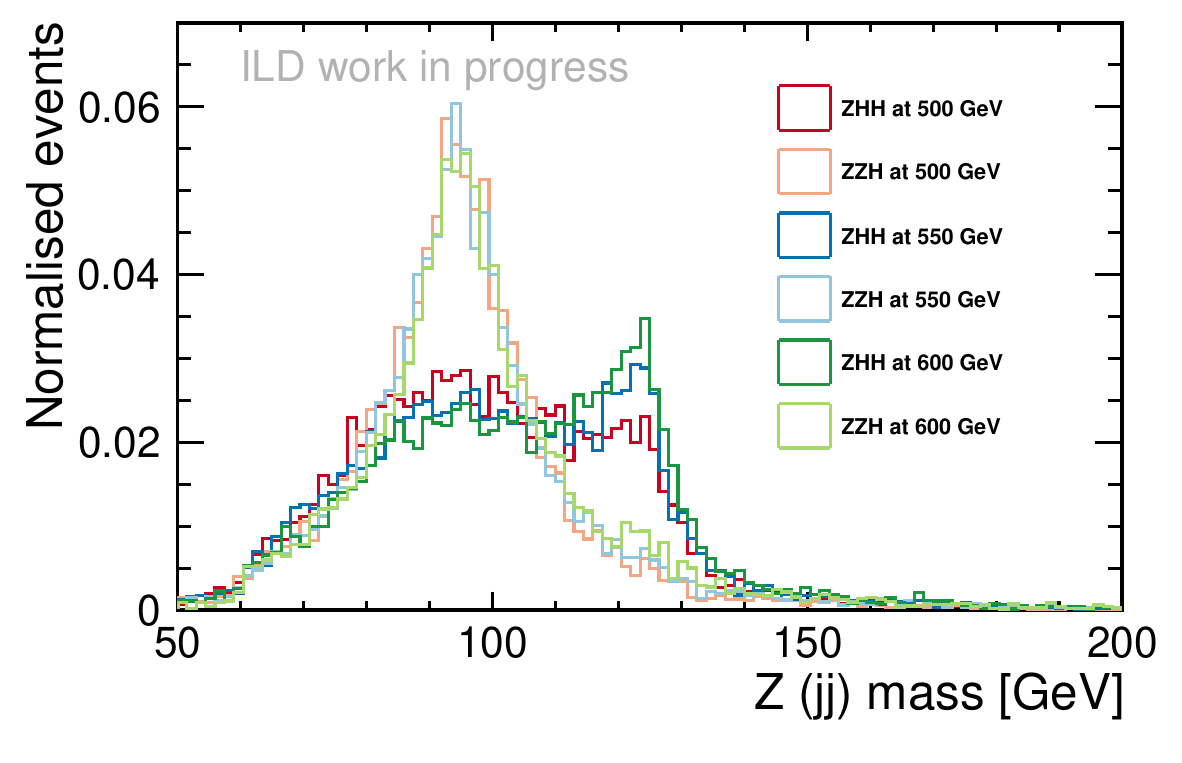}
  \caption{$ZZH$ hypothesis \label{fig:jetpairingZZH}}
\end{subfigure}
\caption{\label{fig:jetpairing} Di-jet masses using the pairing according to a $ZHH$ hypothesis and a $ZZH$ hypothesis, comparing $ZHH$ and $ZZH$ events at $500$, $550$ and $600$\,GeV.}
\end{figure}

The last potential advantage of increasing $E_{CM}$ concerns kinematic separation, for which kinematic fitting can be utilized. 
For the fit to perform well, it is crucial to have a correct error parametrisation of the fit object like the jets. A new tool, called ErrorFlow~\cite{radkhorrami2021kinematic}, that did not exist 10 years ago, parametrises multiple sources of uncertainties for the individual jets, including the detector resolution and particle confusion in the Particle Flow Algorithm~\cite{marshall2013pandora}, and most recently corrects for missing neutrinos within the jets. 
Kinematic fitting is used in the $ZHH$ analysis to perform hypothesis testing to help separate $ZHH$ events from $ZZH$ events, here in the channel where one $Z$ boson decays leptonically, and the other two bosons decay hadronically. Both hypotheses impose 4-momentum conservation where the $ZHH$ hypothesis constrains the di-jet systems to fit with two Higgs masses while the $ZZH$ hypothesis constrains the di-jet systems to fit with a Higgs mass and a $Z$ mass. More details can be found in~\cite{torndal2023higgs}. Figure~\ref{fig:jetpairing} shows the di-jet mass distributions using the jet pairing found from each hypothesis. With the $ZHH$ hypothesis, the di-jet mass distributions for $ZHH$ events are sharper at higher $E_{CM}$, which is also slightly the case for the $ZZH$ events, and with the $ZZH$ hypothesis, more $ZHH$ events remain at the Higgs mass at $550$ and $600$\,GeV compared to $500$\,GeV i.e.\ choosing the wrong jet pairing becomes less likely.

\section{The Higgs self-coupling Beyond the SM}
\begin{figure}
    \centering
    \includegraphics[width=0.6\textwidth]{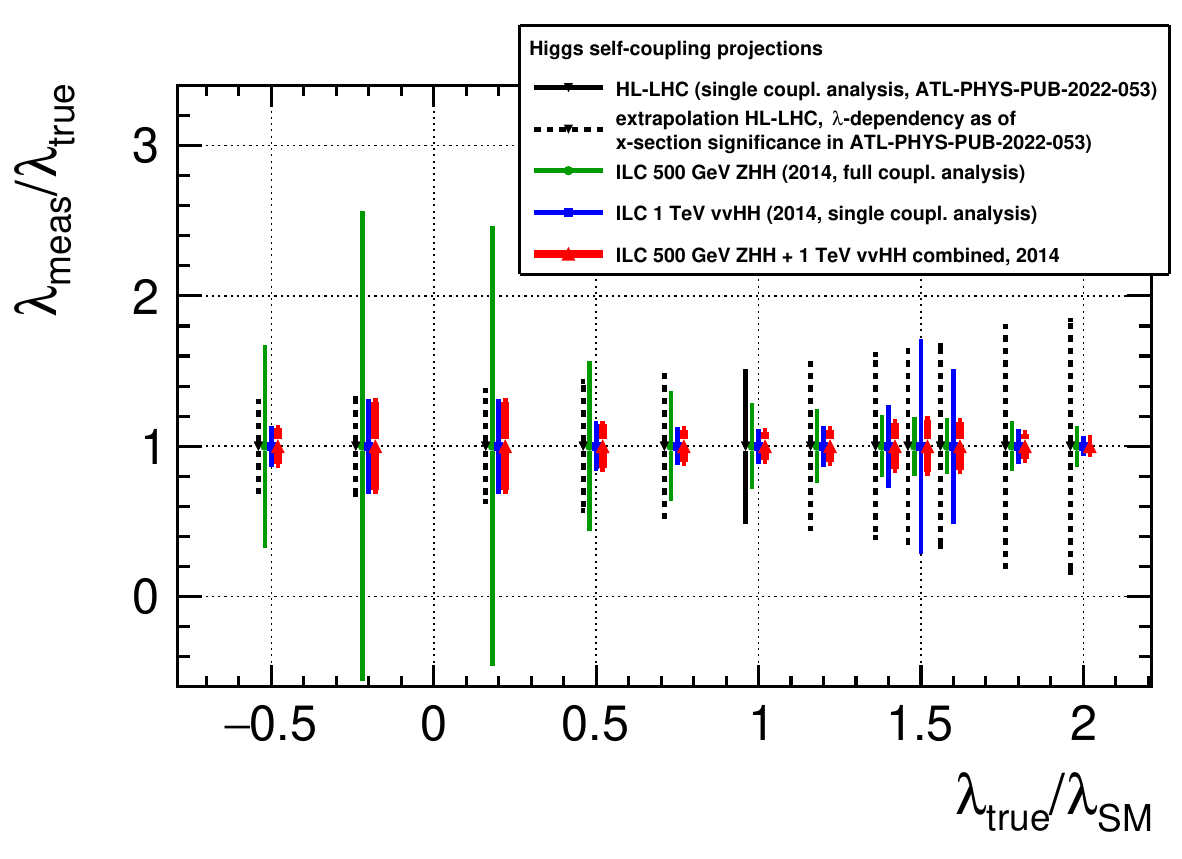}
    \caption{Comparison of various state-of-the-art projections for the Higgs self-coupling measurements at HL-LHC and ILC. The HL-LHC projections are based on~\cite{ATLAS:2022faz}, assuming that the self-coupling measurement precision will change with the actual value of $\lambda$ similarly to the cross-section significance. The ILC projections are based on~\cite{Duerig:310520,LC-REP-2013-025} from 2014, and significant improvements will be expected from the ongoing re-analysis described in these proceedings. Note that only for the extraction of $\lambda$ from $ZHH$ at $500$\,GeV the shown values correspond to a global SMEFT extraction~\cite{Barklow:2017awn}, while all other projections are single-parameter fits.}
    \label{fig:precisionforlambda}
\end{figure}

 It is important to remember that the precisions quoted in Sec.~\ref{sec:lambdameasurement} are only valid when $\lambda$ equals its SM value. However the achievable precision changes significantly with the value of $\lambda$, as it impacts both the size of the cross-section as well as its sensitivity to $\lambda$. Figure~\ref{fig:precisionforlambda} shows the expected precision as a function of $\lambda$. At HL-LHC, the precision improves with lower values of $\lambda$ but degrades as $\lambda$ increases. In $e^+e^-$ collisions, $ZHH$ measurements result in superior precision for higher values of $\lambda$,  while for lower values of $\lambda$ the $\nu\nu HH$ measurement dominates. Thus the two measurements are complementary and combining them ensures 10-15\% precision not only for $\lambda = \lambda_{SM}$, but also for $\lambda > \lambda_{SM}$ as required by electroweak baryogenesis, and at least 30\% precision for \textit{any} value of $\lambda$. 

\section{Conclusion}
While circular $e^+e^-$ colliders are restricted to constraining the triple-Higgs coupling from loop contributions to single Higgs boson processes, future linear $e^+e^-$ colliders such as the ILC, CLIC or C$^3$ offer tree-level sensitivity to the Higgs self-coupling through the measurement of the cross-section for Higgs pair production. In particular ILC and C$^3$ plan to operate at energies of $500$ or $550$\,GeV, respectively, in order to measure the rate of $e^+e^- \to ZHH$. Since the last studies of this proccess for ILC were performed, a lot of progress has been made on improving reconstruction techniques which is foreseen to get the precision of a $\lambda_{SM}$ measurement to better than 20\%. Aspects concerning jet clustering, flavor tagging and kinematic reconstruction have been covered in this contribution focusing both on improvements in the reconstruction tools and impact of the choice of centre-of-mass energy in the range of $500$ to $600$\,GeV. We also stress that is is important to not restrict the scientific discussion to the SM value of $\lambda$, but to study the whole possible range of values, in particular those motivated by electroweak baryogenesis. Only the combination of di-Higgs production measurements from double-Higgs strahlung and double Higgs from $WW$ fusion can ensure coverage of the whole relevant BSM range.

\section{Acknowledgements}
We would like to thank the LCC generator working group and the ILD software working group
for providing the simulation and reconstruction tools and producing the Monte Carlo samples used
in this study. This work has benefited from computing services provided by the ILC Virtual Organ-
isation, supported by the national resource providers of the EGI Federation and the Open Science
GRID. In this study we widely used the National Analysis Facility (NAF) and the Grid computational 
resources operated at Deutsches Elektronen-Synchrotron (DESY), Hamburg, Germany. 
We thankfully acknowledge the support by the Deutsche Forschungsgemeinschaft
(DFG, German Research Foundation) under Germany's Excellence Strategy EXC 2121 "Quantum
Universe" 390833306.

% add references
\printbibliography[title=References]

\end{document}